\documentclass[aps, prl, twocolumn, superscriptaddress]{revtex4}
\usepackage{graphicx}
\usepackage{array}
\usepackage{amsmath,amssymb}
\usepackage{mathrsfs}
\usepackage[dvipsnames,usenames]{color}
\usepackage{float}

\begin{document}
\title{Strongly correlated electrons in superconducting islands with fluctuating Cooper pairs}
\author{Tie-Feng Fang}
\email[]{fangtiefeng@lzu.edu.cn}
\affiliation{School of Sciences, Nantong University, Nantong 226019, China}
\author{Ai-Min Guo}
\affiliation{Hunan Key Laboratory for Super-microstructure and Ultrafast Process, School of Physics and Electronics,
Central South University, Changsha 410083, China}
\author{Qing-Feng Sun}
\email[]{sunqf@pku.edu.cn}
\affiliation{International Center for Quantum Materials, School of Physics, Peking University, Beijing 100871, China}
\affiliation{Collaborative Innovation Center of Quantum Matter, Beijing 100871, China}
\affiliation{CAS Center for Excellence in Topological Quantum Computation, University of Chinese Academy of Sciences, Beijing 100190, China}
\date{\today}
\begin{abstract}
We present a particle-number conserving theory for many-body effects in mesoscopic superconducting islands connected to normal electrodes, which explicitly includes quantum fluctuations of Cooper pairs in the condensate. Beyond previous BCS mean-field descriptions, our theory can precisely treat the pairing and Coulomb interactions over an unprecedentedly broad range of parameters by using the numerical renormalization group method. On increasing the ratio of pairing to Coulomb interactions, the low-energy physics of the system evolves from the spin Kondo to mixed valence regimes and eventually reaches an anisotropic charge Kondo phase, while a crossover from $1e$- to $2e$-periodic Coulomb blockade of transport is revealed at high temperatures. For weak pairing, the superconducting condensate is frozen in the local spin-flip processes but fluctuates in the virtual excitations, yielding an enhanced spin Kondo temperature. For strong pairing, massive fluctuations of Cooper pairs are crucial for establishing charge Kondo correlations whose Kondo temperature rapidly decreases with the pairing interaction. Surprisingly, a charge-exchange induced local field may occur even at the charge degenerate point, thereby destroying the charge Kondo effect. These are demonstrated in the spectral and transport properties of the island.
\end{abstract}
\maketitle

\section{I. Introduction}
Resolving the interplay of different many-body correlations is an exciting challenge in strongly correlated systems. For example, electrons in hybrid nanostructures combining quantum dots (QDs) with bulk superconducting (SC) baths in various geometries \cite{Franceschi2010, Martin-Rodero2011, Meden2019} are strongly correlated. After two decades of extensive studies, these hybrid QD devices have been shown to exhibit a good deal of remarkable emergent phenomena \cite{Siano2004, Bauer2007, Sand-Jespersen2007, Deacon2010, Franke2011, Luitz2012, Kim2013, Oguri2013, Lee2014, Hatter2015, Zitko2015, Delagrange2015, Hatter2017, Cao2017, Lee2017, Fang2017, Fang2018, Saldana2018, Hata2018, Wojcik2019, Lee2019, Su2020, Corral2020, Lim2020, Zalom2021b} due to the competition of SC correlations, Coulomb blockade (CB) \cite{Grabert1992}, and Kondo screening \cite{Hewson1993}. In these studies, the dynamics of the bulk SC condensate is suppressed. The understanding of the competing physics, based on the BCS mean-field theory \cite{Bardeen1957} of superconductivity that violates the number conservation, is excellent both theoretically and experimentally.

Mesoscopic superconductors are, however, not described by the mean-field models, where significant quantum fluctuations and finite charging energy in the SC condensate can have observable consequences \cite{Delft2001, Pavesic2021}. In particular, small SC islands embedded between bulk reservoirs constitute a distinctive class of quantum impurity systems, having the potential to generate unusual electron correlations by SC pairings at the mesoscopic scale. Such SC QD devices were fabricated in early experiments on mesoscopic SC grains \cite{Lafarge1993a, Lafarge1993b, Eiles1993, Hergenrother1994, Hergenrother1995} that demonstrated a prominent CB of Andreev current with a period of two electrons ($2e$) due to the energy asymmetry between even and odd occupancies of the island. Tremendous experimental efforts \cite{Higginbotham2015, Albrecht2016, Albrecht2017, Farrell2018, Shen2018, Vaitiekenas2018, Vaitiekenas2020, Whiticar2020, Carrad2020, Pendharkar2021, Kanne2021, Shen2021} have recently been devoted to examining this $2e$-periodic phenomenon in semiconductor nanowire QDs with proximity-induced mesoscopic superconductivity, as a premise for realizing Majorana physics \cite{Lutchyn2018, Prada2020}. Indeed, the $2e$ periodicity of CB is now equally an experimental hallmark for transport through SC QDs, distinct from the usual single-electron ($1e$) CB in normal QDs. There are also very perceptive theories that address SC QDs in the CB regime \cite{Averin1992, Hekking1993, Houzet2005, Heck2016} and predict the charge Kondo effect \cite{Garate2011, Lutchyn2016, Pustilnik2017, Papaj2019} arising from massive charge fluctuations by $2e$ in the islands. Yet most of them rely on the BCS mean-field approximation or Bogoliubov-de Gennes approach, thereby underestimating the fluctuations of Cooper pairs. Moreover, the calculation of the island ground-state energy with definite electron numbers seems empirical, since its microscopic derivation from the mean-field models violating the number conservation does not exist \cite{Averin1992, Hekking1993, Heck2016, Garate2011}. The theoretical results are thus valid only in limited parameter ranges. Given the potential significance of related experimental observations, it is important to build a more quantitative understanding for SC QDs. Rigorous theoretical insight on strongly correlated effects arising from the SC pairing and Coulomb interactions at the mesoscopic scale is still lacking even without involving topological aspects.

In this paper, we introduce a number-conserving model for SC QDs, which is minimal but fully involves quantum fluctuations of Cooper pairs in the SC condensate. By using the numerical renormalization group (NRG) method, our theory gives a consistent and quantitative account for electron correlations originating from the interplay of pairing and Coulomb interactions. Characteristic spectral and transport features, obeying the Friedel sum rule due to the number conservation, are reported over the full range of interest beyond the scope of BCS mean-field description, from the CB to Kondo effects, from the spin-Kondo to charge-Kondo regimes, and from the $1e$- to $2e$-periodic CB oscillations. Remarkably, we find a charge-exchange field arising at the charge degeneracy and acting as an effective magnetic field in the charge Kondo regime, which has no counterpart in previous theories.

The remainder of the paper is organized as follows. Section II introduces the microscopic model for SC islands coupled to two normal leads and obtains the eigenenergies and eigenstates of isolated islands. The low-energy effective models derived by the Brillouin-Wigner perturbation theory are discussed in Sec.\,III, where the basic scenarios of the $1e$- and $2e$-periodic CB and the spin and charge Kondo physics are demonstrated. Section IV characterizes the resulting spectral density and conductance by using the NRG method. Section V is devoted to a conclusion. Three appendices include full derivation details of the spectra of isolated islands (Appendix A), the Brillouin-Wigner perturbation theory (Appendix B), and the spin and charge Kondo Hamiltonians (Appendix C).

\section{II. Microscopic model}
Specifically, the system we study consists of a small SC island coupled to two normal electrodes, modeled by the Hamiltonian:
$H=H_d+\sum_{\alpha=L,\,R}(H_\alpha+H_{T\alpha})$. The central ingredient is the number-conserving Hamiltonian $H_d$ for the island, which reads
\begin{equation}
H_d=\sum_{\sigma}\varepsilon_0d^\dagger_\sigma d_\sigma+(\Delta d^\dagger_\uparrow d^\dagger_\downarrow e^{-i\hat\phi}+\textrm{h.c.})+\frac{E_c}{2}(\hat N-V_g)^2.
\end{equation}
Here $d^\dagger_\sigma$ creates an electron of spin $\sigma\hspace{-1mm}=\,\,\uparrow,\downarrow$ in the normal energy level $\varepsilon_0$. Its occupation state is denoted by $|\lambda;\,d\rangle$, $\lambda=0,\,\uparrow,\,\downarrow,\,\uparrow\downarrow$. The second term in $H_d$ describes the number-conserving SC pairing, where $\Delta$ is the pairing strength \cite{Note1} and the operator $e^{-i\hat\phi}$, as originally introduced in Refs.\,\cite{Altland2013, Michaeli2017, Sela2017}, annihilates a Cooper pair in the SC condensate. More specifically, the condensate state, $|m;\,s\rangle$, $m\in\mathbb{Z}$, is the eigenstate of the number operator $\hat{N}_p$ of Cooper pairs, $\hat N_p|m;\,s\rangle=m|m;\,s\rangle$. $\hat N_p$ and the phase $\hat\phi$ obey $\big[\hat\phi,\,\hat N_p\big]=i$, yielding $e^{\pm i\hat\phi}|m;\,s\rangle=|m\pm1;\,s\rangle$. The last term in $H_d$ represents the electrostatic energy, where $\hat N=\sum_{\sigma}d^\dagger_\sigma d_\sigma+2\hat N_p$ is the total number of electrons in the island, $E_c$ the Coulomb (charging) energy, and $V_g$ the dimensionless gate voltage. $H_\alpha=\sum_{k,\,\sigma}\varepsilon_kC^\dagger_{k\sigma\alpha}C_{k\sigma\alpha}$ models the left ($L$) and right ($R$) normal leads, with $C^\dagger_{k\sigma\alpha}$ the creation operator in lead $\alpha$. $H_{T\alpha}=\sum_{k,\sigma}t_\alpha(C^\dagger_{k\sigma\alpha}d_\sigma+\text{h.c.})$ is the dot-lead tunneling characterized by the amplitude $t_\alpha$. It is convenient to define a tunneling rate $\Gamma_\alpha\equiv\pi\rho t^2_\alpha$, with $\rho$ the density of lead states.

Our model Hamiltonian (1) describes the SC-QD device consisting of a mesoscopic SC grain \cite{Lafarge1993a, Lafarge1993b, Eiles1993, Hergenrother1994, Hergenrother1995} or a short proximitized nanowire segment \cite{Higginbotham2015, Albrecht2016, Albrecht2017, Farrell2018, Shen2018, Vaitiekenas2018, Vaitiekenas2020, Whiticar2020, Carrad2020, Pendharkar2021, Kanne2021, Shen2021}. While the former can host SC correlations on its own, the latter is superconducting by proximity coupling to a mesoscopic superconductor. In either case, the whole device is floating and its charge is tuned by a capacitively coupled gate voltage, giving rise to the finite charging energy of Cooper pairs. The charge conservation of the pairing interaction is thus necessary in order to properly describe the charge fluctuations. Moreover, we consider only one spin-degenerate normal state closest to the Fermi energy in SC islands, which as shown later is sufficient to capture the prominent $2e$ periodicity observed in the experiments \cite{Lafarge1993a, Lafarge1993b, Eiles1993, Hergenrother1994, Hergenrother1995, Higginbotham2015, Albrecht2016, Albrecht2017, Farrell2018, Shen2018, Vaitiekenas2018, Vaitiekenas2020, Whiticar2020, Carrad2020, Pendharkar2021, Kanne2021, Shen2021}.

If one removes the operators $e^{\pm \hat\phi}$ and $\hat N_p$ from Eq.\,(1), the pairing term will violate the number conservation and the model will reduce to the extensively studied system of a QD in the SC atomic limit \cite{Martin-Rodero2011, Meden2019, Bauer2007, Fang2017, Fang2018, Domanski2016, Domanski2017, Zalom2021}. This limit can be achieved by proximity coupling the dot to an additional SC electrode with an infinite gap and no charging energy. As shown below, the properties of our SC QDs modelled by Eq.\,(1) that explicitly involves the degrees of freedom of Cooper pairs are radically different from those of a QD in the SC atomic limit.

The Hilbert space of an isolated SC QD described by Eq.\,(1) is spanned by the direct products $|\lambda,\,m\rangle\equiv|\lambda;\,d\rangle\otimes|m;\,s\rangle$. In this basis, because of the electron-number conservation $\big[H_d,\,\hat N\big]=0$, the eigenenergy $E^{\sigma,\nu}_N(V_g)$ and eigenstate $\big|\Phi^{\sigma,\nu}_N\big\rangle$ of $H_d$ depend on the eigenvalue $N$ of $\hat N$ (see Appendix A for details): If $N$ is odd,
\begin{equation}
E^\sigma_N(V_g)=\varepsilon_0+\tfrac{1}{2}E_c(N-V_g)^2\equiv E^0_{N}(V_g),\\
\end{equation}
and $\big|\Phi^{\sigma}_N\big\rangle=\big|\sigma,\,\tfrac{N-1}{2}\big\rangle$; If $N$ is even, one has ($\nu=\pm$)
\begin{equation}
E^\nu_N(V_g)=\varepsilon_0+\tfrac{1}{2}E_c(N-V_g)^2+\nu\sqrt{\Delta^2+\varepsilon^2_0},\\
\end{equation}
and $\big|\Phi^{\nu}_N\big\rangle=l_{0\nu}\big|0,\,\tfrac{N}{2}\big\rangle+l_{2\nu}\left\vert\uparrow
\downarrow,\,\tfrac{N-2}{2}\right\rangle$, with $l_{0\nu}=c_\nu l_{2\nu}$, $l_{2\nu}=1/\sqrt{1+c^2_\nu}$, $c_\nu=\nu\sqrt{1+\eta^2}-\eta$, and $\eta=\varepsilon_0/\Delta$. Apparently, the even-parity minimal energy $E^-_N(V_g=N)$ is lower by an amount $\sqrt{\Delta^2+\varepsilon^2_0}$ than the odd-parity one $E^\sigma_N(V_g=N)$. While similar formulas for $E^-_N(V_g)$ and $E^\sigma
_N(V_g)$ have been empirically imposed in the literature \cite{Lafarge1993a, Lafarge1993b, Hergenrother1994, Eiles1993, Averin1992, Heck2016, Garate2011}, our theory unambiguously reveals their microscopic origins. Moreover, a new excited state (i.e., $\big|\Phi^{+}_N\big\rangle$) absent in previous theories is created. The energy spectra calculated by Eqs.\,(2) and (3) are illustrated in Fig.\,1, based on which the low-energy physics of our system can be determined.

\begin{figure}
\hspace{-6.5mm}
\includegraphics[width=1.0\columnwidth]{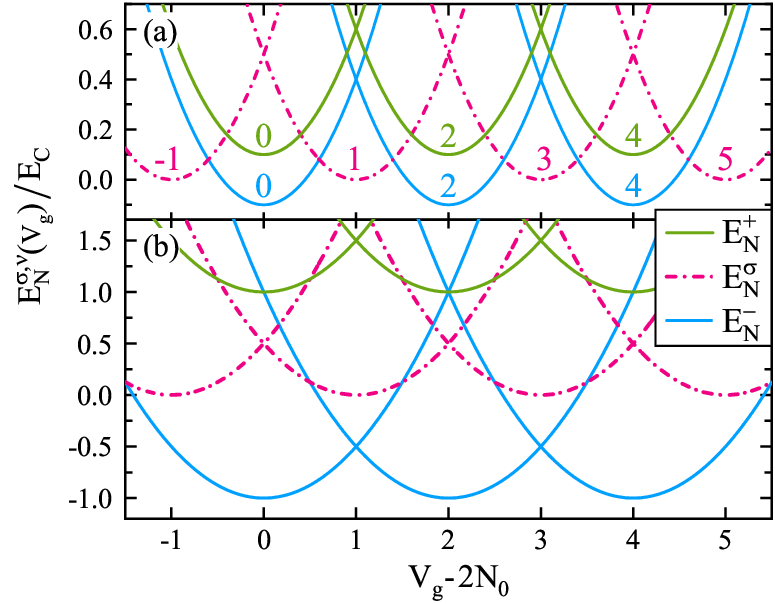}
\caption{Eigenenergy $E^{\sigma,\nu}_N$ of isolated SC QDs as a function of gate voltage $V_g$ for different numbers $N$ of dot electrons, in the regimes of weak [(a) $\Delta/\Delta_c=0.2$] and strong [(b) $\Delta/\Delta_c=2$] SC correlations. Numbers in (a) indicate the value of $N-2N_0$ of each parabola. Here, the $d$ level is set at $\varepsilon_0=0$.}
\end{figure}

\section{III. low-energy effective Kondo Hamiltonians}
For weak SC pairing $\sqrt{\Delta^2+\varepsilon_0^2}<E_c/2\equiv \Delta_c$ [Fig.\,1(a)], the ground-state degeneracy of $N$ and $N+1$ electrons occupying the island occurs at the gate voltage $V_g=2N_0+1\pm K\equiv V_{g\pm}$, with $N_0$ an arbitrary integer and $K=\frac{1}{2}-\sqrt{\Delta^2+\varepsilon_0^2}/E_c$. This can lead to the $1e$-periodic CB of single-electron tunneling \cite{Grabert1992}. In odd-parity CB valleys ($V_{g-}<V_g<V_{g+}$), virtual excitations from the spin doublet $\big|\Phi^\sigma_{2N_0+1}\big\rangle$ to the even-parity states can constitute the spin Kondo effect. Using the Brillouin-Wigner perturbation theory \cite{Fradkin2013} to consider virtual processes up to second order in $t_\alpha$, we derive the effective spin-exchange interaction of SC QDs (see Appendixes B and C for details),
\begin{eqnarray}
H^{\text{s}}_{\text{eff}}&=&\sum_{k,\,k^{\prime }}\left( J_{0}+J_{2}\right) \big[
C_{k\uparrow}^{\dag }C_{k^{\prime }\downarrow
}S^{-}+C_{k^{\prime }\downarrow }^{\dag }C_{k\uparrow
}S^{+}\nonumber\\
&&+\,\,\big( C_{k\uparrow}^{\dag }C_{k^{\prime }\uparrow
}-C_{k\downarrow}^{\dag }C_{k^{\prime }\downarrow
}\big) S^{z}
\big] \left\vert N_{0};s\right\rangle \left\langle N_{0};s\right\vert,
\end{eqnarray}
with $C_{k\sigma}=\sum_\alpha\frac{t_\alpha}{\sqrt{t^2_L+t^2_R}}C_{k\sigma\alpha}$, and the spin operators of the QD, $S^+=(S^-)^\dagger=d^\dagger_\uparrow d_\downarrow$, and $S^z=\frac{1}{2}(d^\dagger_\uparrow d_\uparrow-d^\dagger_\downarrow d_\downarrow)$. The exchange coupling $J_x$ ($x=0,\,2$) reads
\begin{equation}
J_{x}=\sum_{\nu}\frac{(\delta_{x2}l_{0\bar\nu}^{2}+\delta_{x0}l_{2\bar\nu}^{2})(t_L^2+t^2_R)} {E^{\nu}_{2N_0+x}(V_g)-E^0_{2N_0+1}(V_g)}.
\end{equation}
$H^{\text{s}}_{\text{eff}}$ describes the spin Kondo effect, resulting from spin fluctuations in the $d$ level and freezing $N_0$ Cooper pairs in the SC condensate. But in virtual states $\big|\Phi^\nu_{2N_0+x}\big\rangle$ the condensate does fluctuate between $N_0$ and $N_0\pm1$ Cooper pairs. The second-order scaling \cite{Hewson1993} yields the Kondo temperature, $T_K^s\sim\exp{[\frac{-1}{2\rho(J_0+J_2)}]}$. At $\Delta=0$, the coupling $J_x$ and hence the Kondo temperature $T^s_K$ restore the values of normal QDs \cite{Hewson1993}. Note that $T_K^s$ is a monotonically increasing function of the pairing $\Delta$.

A similar spin Kondo effect was previously found in the system of a QD in the SC atomic limit \cite{Fang2018, Domanski2016, Domanski2017, Zalom2021}. Its Kondo temperature also increases with $\Delta$, qualitatively agreeing with ours. However, $T_K^s$ of our SC QDs is still quantitatively different from the Kondo temperature of that system, as a result of our number-conserving treatment of Cooper pairs. A more important difference in the spin Kondo effect between the two systems will be demonstrated by numerical results presented in the next section.

For strong pairing $\sqrt{\Delta^2+\varepsilon_0^2}>\Delta_c$ [Fig.\,1(b)], the ground state of the island at $V_g=2N_0+1$ has the degeneracy between adjacent even-parity states $\big|\Phi^-_{2N_0}\big\rangle$ and $\big|\Phi^-_{2N_0+2}\big\rangle$, leading to the $2e$-periodic CB of Andreev reflection \cite{Eiles1993, Hergenrother1994, Hergenrother1995}. At the degeneracy point, charge Kondo correlations may also occur due to virtual excitations from this ground-state charge doublet to the odd-parity subspace. The second-order perturbation theory yields a pseudospin-exchange Hamiltonian (see Appendixes B and C for details),
\begin{eqnarray}
H^{\text{c}}_{\text{eff}}&=&\tilde{B}Q^{z}+\sum_{k,\,k^\prime}J_{\pm}\big(C_{k\uparrow}^{\dag }C_{k^{\prime }\downarrow}^{\dag }Q^{-}
+C_{k^{\prime }\downarrow}C_{k\uparrow}Q^{+}\big)\nonumber\\
&&+\sum_{k,\,k^\prime,\,\sigma}J_{z}\Big(C_{k\sigma}^{\dag }C_{k^{\prime }\sigma}
-\frac{1}{2}\delta_{kk^{\prime }}\Big) Q^{z},
\end{eqnarray}
where $Q^+=(Q^-)^\dagger=\big\vert\Phi^-_{2N_{0}+2}\big\rangle\big\langle\Phi^-_{2N_{0}}
\big\vert$ and $2Q^z=\big\vert\Phi^-_{2N_{0}+2}\big\rangle\big\langle\Phi^-_{2N_{0}+2}
\big\vert-\big\vert\Phi^-_{2N_{0}}\big\rangle\big\langle\Phi^-_{2N_{0}}
\big\vert$ are the pseudospin of the island, and $\tilde{B}=(l_{0-}^2-l_{2-}^2)\sum_{k,\,k^\prime}J_z\delta_{kk^\prime}$. The transverse exchange coupling
$J_{\pm }=J_1$ differs from the longitudinal one
$J_{z}=\frac{l_{2-}}{2l_{0-}}J_{-1}
-\big(\frac{l_{2-}}{2l_{0-}}+\frac{l_{0-}}{2l_{2-}}\big)J_1
+\frac{l_{0-}}{2l_{2-}}J_3$. Here $J_y$ ($y=-1,\,1,\,3$) is given by
\begin{equation}
J_y=\frac{2l_{0-}l_{2-}(t^2_L+t^2_R)}{E^-_{02}-E^0_{2N_0+y}(V_g=2N_0+1)},
\end{equation}
with $E^-_{02}=\varepsilon_0+\Delta_c-\sqrt{\Delta^2+\varepsilon_0^2}$ the energy of the ground-state doublet. $H^{\text{c}}_{\text{eff}}$ describes a charge Kondo effect arising from charge fluctuations between spinless states $|\Phi^-_{2N_0}\rangle$ and $|\Phi^-_{2N_0+2}\rangle$, subject to a pseudo-magnetic field $\tilde{B}$. The field $\tilde{B}$ is induced by charge exchange processes between the island and leads, which is finite only if both the coupling $t_\alpha$ and $d$-level $\varepsilon_0$ are nonzero. Such an exchange field, taking place at the charge degeneracy point $V_g=2N_0+1$, has never been revealed in previous BCS mean-field studies of the charge Kondo effect in SC islands \cite{Garate2011, Lutchyn2016, Pustilnik2017, Papaj2019}. Moreover, in our model, virtual transitions violate pseudospin rotational invariance, yielding a Kondo temperature \cite{Zitko2008, Schoeller2009, Yoo2014}, $T_K^c\sim\exp{(\frac{-\eta}{2\rho J_z})}$, with $\eta=\frac{\arctan\gamma}{\gamma}$ and $\gamma=\sqrt{(J_\pm/J_z)^2-1}$. At $\varepsilon_0=0$, the anisotropy reduces to $J_\pm/J_z=\frac{1}{4}(\frac{\Delta}{\Delta_c}+3)$, increasing linearly with the SC pairing, while the Kondo temperature $T_K^c$ decreases rapidly.

The above physical scenario of our SC QDs in the strong-pairing regime is in stark contrast to those of a QD in the SC atomic limit. According to the results known in the literature \cite{Martin-Rodero2011, Meden2019, Bauer2007, Fang2017, Fang2018, Domanski2016, Domanski2017, Zalom2021}, for strong pairing, the ground state of a QD in the SC atomic limit is always a BCS singlet and there is no $2e$ periodicity and no charge Kondo effect at all when normal leads are coupled.

In the intermediate regime $\sqrt{\Delta^2+\varepsilon^2_0}\backsimeq\Delta_c$, the states of isolated SC QDs, $\big|\Phi^-_{2N_0}\big\rangle$, $\big|\Phi^\sigma_{2N_0+1}\big\rangle$, and $\big|\Phi^-_{2N_0+2}\big\rangle$, are degenerate at the gate voltage $V_g=2N_0+1$, constituting the ground state with a four-fold degeneracy. The low-energy physics at this point is dominated by first-order tunneling processes and the system is in the mixed valence situation. The effective model does not describe the higher symmetric Kondo effects as found in similar systems of a QD coupled to a Coulomb box and normal leads \cite{Mitchell2021, Hur2004, Borda2003}.

\section{IV. NRG Results and discussion}
We now turn to characterize the consequences of the above arguments in local spectral density and transport properties of SC QDs. The full Hamiltonian $H$ has been solved by using the full density-matrix NRG method \cite{Wilson1975, Krishna-murthy1980, Bulla2008, Anders2005, Peters2006, Weichselbaum2007, Fang2015} modified to include the degrees of freedom of Cooper pairs. For each value of the gate voltage $V_g$, the total electron number $N$ in the island is set to fluctuate in the interval $[V_g]-F\leqslant N\leqslant[V_g]+F$. We take $F=50$ in order to converge the NRG calculations for the set of model parameters: $N_0=100$, $\Gamma_L=\Gamma_R$, $\Gamma\equiv2\Gamma_L=0.007D$, $\varepsilon_0=0$ (unless stated otherwise), the half Coulomb energy $\Delta_c=0.06D$, and half bandwidth $D=1$. Smaller values of $F$ would result in the violations of relevant sum rules for the spectral function and the periodicity with the gate voltage. Our NRG calculations were carried out by using a discretization parameter $\Lambda=1.8$ and retaining 2400 states per iteration. The data for spectral density is smoothened based on the log-Gaussian kernel with a broadening parameter $\alpha=0.54$, while the conductance obtained by summing over discrete data is more accurate than the spectral density. Numerical results are presented below.

\begin{figure}
\hspace{-3mm}
\includegraphics[width=1.0\columnwidth]{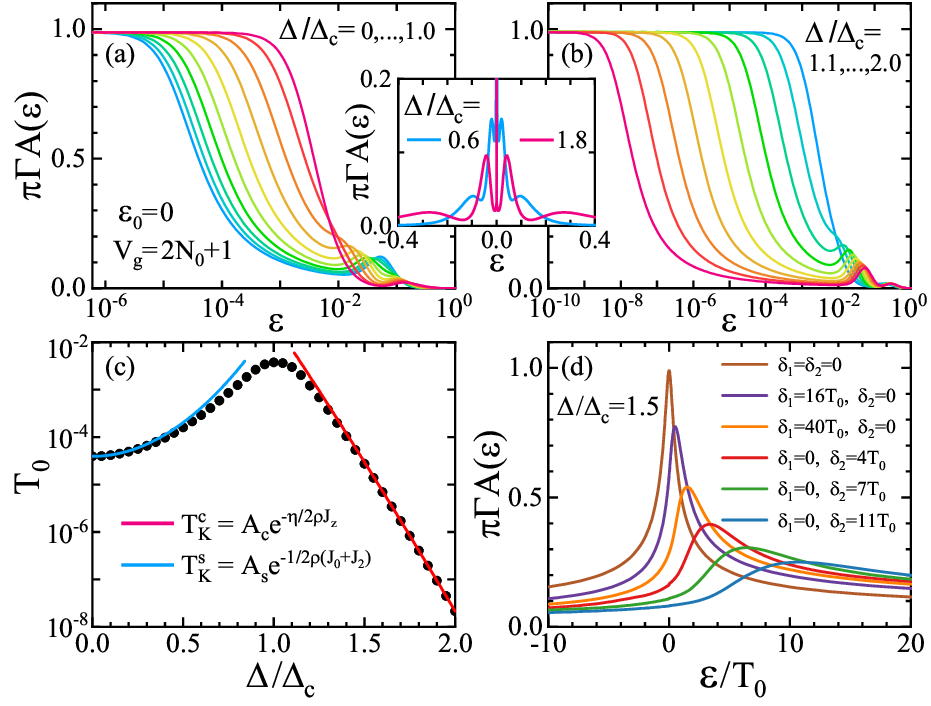}
\caption{Zero-temperature spectral properties of $d$ electrons in SC islands. (a) Spectral density $A(\varepsilon)$ vs positive energy $\varepsilon>0$, for different SC pairing, $\Delta/\Delta_c=0$(blue), $0.2$, $0.3$, $0.4$, $0.5$, $0.6$, $0.7$, $0.8$, $0.9$, and $1.0$(red). (b) Same as (a) but for $\Delta/\Delta_c=1.1$(blue), $1.2$, $1.3$, $1.4$, $1.5$, $1.6$, $1.7$, $1.8$, $1.9$, and $2.0$(red). Inset: $A(\varepsilon)$ in the full range of $\varepsilon$. (c) Half width at half maximum $T_0$ of the central peak in $A(\varepsilon)$ as a function of $\Delta$. Two solid lines are fits by two analytical expressions of Kondo temperatures, with $A_c=0.02686$ and $A_s=0.02741$. (d) Suppression of the charge Kondo resonance by the $d$ level $\varepsilon_0$ ($\delta_1=2\varepsilon_0$) and the gate voltage $V_g$ [$\delta_2=2E_c(2N_0+1-V_g)$] deviating from the symmetric point.}
\end{figure}

Figure 2 presents the $d$-electron spectral density $A(\varepsilon)=-\frac{1}{\pi}\textrm{Im}\langle\langle d_\sigma; d^\dagger_\sigma\rangle\rangle$ of SC QDs at zero temperature. Here $\langle\langle d_\sigma; d^\dagger_\sigma\rangle\rangle$ is the retarded Green's function for $d$ electrons and $A(\varepsilon)$ is spin independent in the absence of external magnetic field. As shown in Figs.\,2(a) and 2(b), the case of gate voltage at odd values $V_g=2N_0+1$, which restores the particle-hole symmetry $A(\varepsilon)=A(-\varepsilon)$ (see inset), yields the most striking spectral features. For zero SC pairing $\Delta=0$, the SC QD reduces to a normal QD and $A(\varepsilon)$ exhibits typical three-peak structure in the spin Kondo effect \cite{Hewson1993}. Upon increasing $\Delta$, the central spin Kondo resonance broadens due to the enhancement of the exchange couplings $J_x$ [Eq.\,(5)], while the Hubbard sidebands split because of the splitting of even-$N$ levels $E_N^+$ and $E_N^-$ at finite $\Delta$. Specifically, the positions of Hubbard sidebands are determined by the energy differences between the ground state $\big|\Phi_{2N_0+1}^\sigma\big\rangle$ and the excited states $\big|\Phi_{2N_0}^\pm\big\rangle$ and $\big|\Phi_{2N_0+2}^\pm\big\rangle$ of the isolated islands. The split Hubbard bands moving toward the Fermi energy eventually merge with the Kondo peak as $\Delta$ approaches the half Coulomb energy $\Delta_c$. In this mixed-valence regime, the four states of the island, $\big|\Phi_{2N_0}^-\big\rangle$, $\big|\Phi_{2N_0+1}^\sigma\big\rangle$, and $\big|\Phi_{2N_0+2}^-\big\rangle$, are degenerate and as a result of the first-order tunneling processes, the spectral density $A(\varepsilon)$ features a broad resonance at $\varepsilon=0$ with a width of the order of the hybridization $\Gamma$. The broad resonance splits again by further increasing the pairing energy excess the half Coulomb energy ($\Delta>\Delta_c$). This drives the SC QD into the charge Kondo regime characterized by a narrowing Kondo resonance at the Fermi energy.

The spectral data in Figs.\,2(a) and 2(b) further demonstrates that the resonance at the Fermi energy satisfies (within a tiny error of NRG) the unitarity condition $\pi\Gamma A(0)=1$ of the Friedel sum rule \cite{Hewson1993} in the particle-hole symmetric case for arbitrary $\Delta$ ranging from the spin-Kondo to mixed-valence and charge-Kondo regimes. Obeying certain Fermi-liquid relation is an important merit of our number-conserving theory for SC islands. This is not the case for the system of a QD in the SC atomic limit, where the unitarity condition is violated \cite{Fang2018, Domanski2016, Domanski2017, Zalom2021} for any nonzero pairing even in the spin-Kondo regime.

While the peak height is always unitary as guaranteed by relevant sum rules, the half width $T_0$ of the central peak in our spectral density $A(\varepsilon)$ of SC islands exhibits a nonmonotonic dependence on the SC pairing $\Delta$ as shown in Fig.\,2(c). Figure 2(c) illustrates discrete $T_0(\Delta)$ as a function of $\Delta$, where fits to the analytical expressions of the spin $T^s_K$ and charge $T_K^c$ Kondo temperatures are also presented. We use $T_K^s=A_s\exp[\frac{-1}{2\rho(J_0+J_2)}]$ and $T_K^c=A_c\exp(\frac{-\eta}{2\rho J_z})$ for the weak ($\Delta<\Delta_c$) and strong ($\Delta>\Delta_c$) superconductivities, respectively. Only the prefactors $A_s$ and $A_c$ are fitting parameters. It can be seen from Fig.\,2(c) that the agreement is excellent, confirming the Kondo nature of the central peak in $A(\varepsilon)$. In the mixed valence region ($\Delta\sim\Delta_c$), charge fluctuations are not Coulomb blockaded because of the dominant first-order tunnelings, giving rise to $T_0(\Delta=\Delta_c)\simeq0.54\Gamma$.

Our charge Kondo effect differs from that in the negative-$U$ Anderson impurity \cite{Taraphder1991, Koch2007, Andergassen2011, Costi2012, Fang2014, Fang2017, Fang2018} where Kondo correlations are attributed to charge fluctuations between zero and double occupancies of the $d$ level. In our SC QDs, the pseudospin represents the two many-body ground states $|\Phi_{2N_0}^-\rangle$ and $|\Phi_{2N_0+2}^-\rangle$ of the island. Although each state contains also the zero-double fluctuations in the $d$ level, such fluctuations do not contribute to the charge Kondo effect. In order to demonstrate this, we plot in Fig.\,2(d) the suppression and displacement of the charge Kondo resonance by shifting the $d$ level $\varepsilon_0$ or the gate voltage $V_g$ from the symmetric point. For nonzero $\varepsilon_0$, although isolated SC islands always remain the charge degeneracy as long as $V_g=2N_0+1$, the exchange field $\tilde{B}$ induced by the dot-lead tunnelings can split the ground-state doublet, resulting in $E_{2N_0+2}^--E_{2N_0}^-=\tilde{B}$ [see Eq.\,(6)]. On the other hand, the energy difference between zero and double occupancies of the $d$ level is $\delta_1=2\varepsilon_0$. Figure 2(d) shows that the position of the suppressed Kondo peak is not at $\varepsilon\simeq\delta_1$. Instead, it is given by $\varepsilon\simeq\tilde{B}\ll\delta_1$. For $V_g$ deviating from $2N_0+1$, one has $E_{2N_0+2}^--E_{2N_0}^-=2E_c(2N_0+1-V_g)\equiv\delta_2$ calculated by Eq.\,(3). The resulting suppressed Kondo peak appears at $\varepsilon\simeq\delta_2$, as also shown in Fig.\,2(d). These observations indicate that our charge Kondo resonance in the $d$-electron spectral density does indeed arise from Cooper-pair fluctuations between the island states $\big|\Phi_{2N_0}^-\big\rangle$ and $\big|\Phi_{2N_0+2}^-\big\rangle$, not from simple fluctuations in the $d$ level.

\begin{figure}
\hspace{0mm}
\includegraphics[width=1.0\columnwidth]{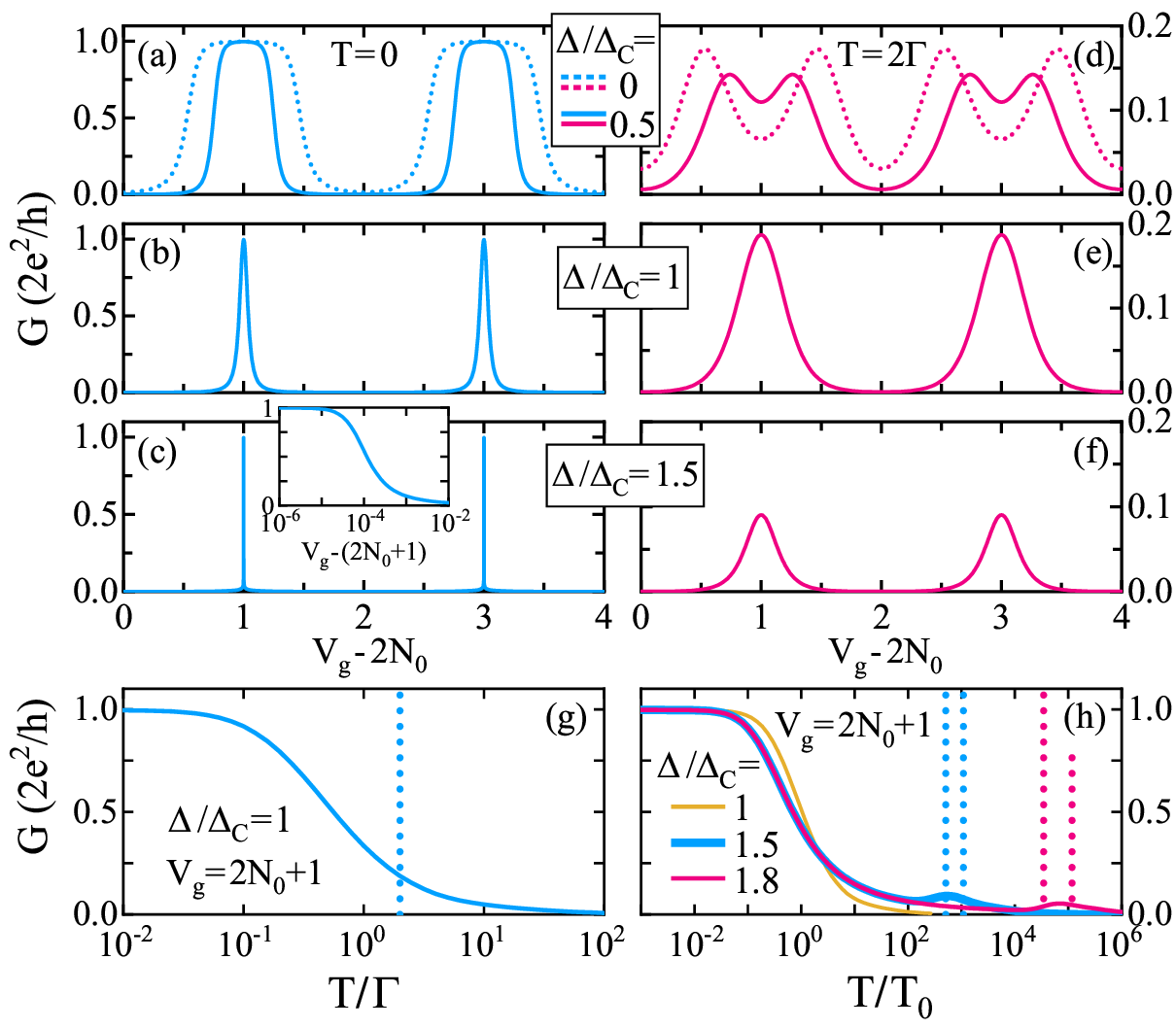}
\caption{(a)-(f) Conductance $G$ as a function of the gate voltage $V_g$ at zero temperature $T=0$ [(a)-(c)] and high temperature $T=2\Gamma$ [(d)-(f)], for several values of the SC pairing $\Delta$. Inset of (c): Zoom into the resonance at $V_g=2N_0+1$. (g) and (h) Conductance as a function of the temperature $T$ scaled respectively by $\Gamma$ and $T_0$ for different $\Delta$. The vertical line in (g) indicates the position of $T=2\Gamma$. In (h), two blue vertical lines mark the positions of $T=2\Gamma$ (left) and $E^*$ (right) for $\Delta/\Delta_c=1.5$, with $E^*$ the excitation energy (see text). Two red vertical lines represent the same but for $\Delta/\Delta_c=1.8$.}
\end{figure}

We present in Fig.\,3 the conductance $G$ through SC QDs in the linear-response regime \cite{Meir1992},
\begin{equation}
G=-\frac{2e^2}{h}\pi\Gamma\int_{-D}^{D} A(\varepsilon)\frac{\partial f(\varepsilon)}{\partial\varepsilon}\,d\varepsilon,
\end{equation}
with $f(\varepsilon)$ the Fermi distribution function, as a function of the gate voltage $V_g$ and the temperature $T$. For weak pairing ($\Delta<\Delta_c$), the zero-temperature conductance is prominent by broad spin-Kondo plateaus of unitary transmission centered at odd values of $V_g$ [Fig.\,3(a)], while at high $T$, the plateaus collapse and the conductance exhibits the $1e$-periodic CB oscillation with peaks occurring at the even-odd degeneracy points [Fig.\,3(d)]. These transport features resemble those of normal QDs \cite{Goldhaber-Gordon1998, Cronenwett1998, Nygard2000, Wiel2000}. The effect of a finite $\Delta$ in this regime at zero (high) temperature is to narrow the width of the Kondo plateaus (odd CB valleys). When the pairing energy increases to the mixed valence region ($\Delta\simeq\Delta_c$), the narrowing at all temperatures leads to a single conductance peak formed at odd values of $V_g$, giving rise to the celebrated $2e$ periodicity of CB [Figs.\,3(b) and 3(e)] \cite{Albrecht2016}. Here the transport at the $2e$ CB peaks is mediated by resonant Andreev and single-electron tunnelings, which at zero temperature can reach the unitary limit due to the large degeneracy of island states discussed above. The zero-temperature width of conductance CB peaks in the gate voltage is $W\simeq\Gamma/2E_c$. Accordingly, the CB peaks at finite $T$ follow a temperature smearing scaled also by the dot-lead coupling $\Gamma$ [Fig.\,3(g)].

For strong pairing ($\Delta>\Delta_c$), the overall conductance feature of SC QDs is still the $2e$-periodic CB oscillation [Figs.\,3(c) and 3(f)], but now the even-even degeneracy at odd values of $V_g$ can support only the second-order Andreev tunnelings. At temperatures below the Kondo scale $T<T_0$, the coherent superposition of massive Andreev processes can constitute the charge Kondo effect, leading to the unitary transmission as $T\rightarrow0$ [Fig.\,3(c)]. The width $W$ of the charge-Kondo CB peaks in the gate voltage is determined by the Kondo temperature, $W\simeq T_0/2E_c$ [inset of Fig.\,3(c)]. As the temperature increases, the conductance peak fades out following a universal Kondo scaling that is distinct from the $T$-smearing in the mixed valence regime [Fig.\,3(h)]. Interestingly, when the temperature rises to the first excitation energy, $E^*\equiv E^0_{2N_0+1}-E_{02}^-$, of the island, a small peak appears in the conductance vs $T$, as shown in Fig.\,3(h). The underlying physics of this phenomenon is the opening of an additional transport channel for the single-electron tunnelings. Since the 2$e$-periodic CB peaks as in Figs.\,3(b), 3(e), and 3(f) are now routinely measured in the experiments on SC islands \cite{Higginbotham2015, Albrecht2016, Albrecht2017, Farrell2018, Shen2018, Vaitiekenas2018, Vaitiekenas2020, Whiticar2020, Carrad2020, Pendharkar2021, Kanne2021, Shen2021}, we believe that at sufficiently low temperature, such experiments can also observe the 2$e$-periodic charge Kondo peaks as in Fig.\,3(c) by fine-tuning the gate voltage.

\section{V. Conclusion}
In summary, we have studied the spectral and transport properties of SC QDs based on a minimal model that conserves the particle number and explicitly takes account of Cooper-pair fluctuations. The merit of the model allows us to accurately treat the SC pairing and Coulomb interactions on an equal footing by using the powerful NRG method. Rigorous results on many-body effects have thus been obtained over the full parameter space from the spin-Kondo to mixed-valence and charge-Kondo regimes. In particular, the number conservation inherent in the theory guarantees that the resulting spectral density obeys the Friedel sum rule, and our theory gives a faithful description of the $2e$-periodic CB phenomena observed in SC islands. For strong SC pairing, we find an unexpected charge-exchange field arising from the dot-lead tunneling, which may suppress the charge Kondo effect even at the charge degenerate point. Our work can serve as a starting point for precisely exploring electron correlations in more complex SC islands, such as proximitized nanowire segments \cite{Lutchyn2018, Prada2020} where the presence of several energy levels, spin-orbit coupling, and topological superconductivity may produce more fascinating emergent phenomena.

\section{Acknowledgments}
We thank H.\,T. Luo for helpful discussions. This work is supported by the National Natural Science Foundation of China (Grants No.\,11874428, No.\,11874187, and No.\,11921005), the National Key Research and Development Program of China (Grant No.\,2017YFA0303301), and the High Performance Computing Center of Central South University.

\begin{widetext}
\begin{appendix}
\section{Appendix A: Isolated superconducting quantum dots}
\setcounter{equation}{0}
\renewcommand{\theequation}{A\arabic{equation}}
The number-conserving Hamiltonian of an isolated superconducting island (quantum dot) reads
\begin{equation}
H_{d}=\sum_{\sigma=\uparrow,\downarrow}\varepsilon _{\sigma}d_{\sigma }^{\dag }d_{\sigma }+\Delta d_{\uparrow
}^{\dag }d_{\downarrow }^{\dag }e^{-i\hat{\phi}}+\Delta e^{i\hat{\phi}%
}d_{\downarrow }d_{\uparrow }+\frac{1}{2}E_{c}\left( \hat{N}-V_{g}\right)
^{2}.
\end{equation}
Since $H_d$ commutes with $\hat{N}$, we can focus on the subspaces characterized by the total electron number $N$. Here, we assume $N>0$ (i.e, excluding the $N=0$ subspace). If $N$ is odd, the basis set of the subspace is $\left\vert \uparrow,\,\frac{N-1}{2}\right\rangle $ and $%
\left\vert \downarrow,\,\frac{N-1}{2}\right\rangle $. Note that
\begin{eqnarray}
H_{d}\left\vert \uparrow,\,\tfrac{N-1}{2}\right\rangle  &=&\left[ \varepsilon _{\uparrow
}+\frac{1}{2}E_c(N-V_{g})^2 \right] \left\vert \uparrow,\,\tfrac{N-1}{2}\right\rangle,  \\
H_{d}\left\vert \downarrow,\,\tfrac{N-1}{2}\right\rangle  &=&\left[ \varepsilon _{\downarrow
}+\frac{1}{2}E_c(N-V_{g})^2 \right] \left\vert \downarrow,\,\tfrac{N-1}{2}\right\rangle.
\end{eqnarray}
The matrix form of $H_{d}$ is
\begin{eqnarray}
H_{d} &=&\begin{pmatrix} \varepsilon _{\uparrow }+\frac{1}{2}E_c(
N-V_{g})^2 & 0 \\ 0 & \varepsilon _{\downarrow }+\frac{1}{2}E_c
(N-V_{g})^2\end{pmatrix}.
\end{eqnarray}
We denote the eigenenergies and eigenstates of the Hamiltonian in this odd subspace as
\begin{eqnarray}
E_N^\sigma(V_g)&=&\varepsilon_\sigma+\frac{1}{2}E_c(N-V_g)^2,\\
|\Phi_N^\sigma\rangle&=&\left|\sigma,\,\tfrac{N-1}{2}\right\rangle.
\end{eqnarray}
In addition, the following formulas for $d_\uparrow$ and $d_\downarrow$ are useful for calculating their matrix elements which will be used in the NRG procedure,
\begin{eqnarray}
d_{\uparrow }\left\vert \uparrow,\,\tfrac{N-1
}{2}\right\rangle=\left\vert 0,\,
\tfrac{N-1}{2}\right\rangle,\hspace{1.5cm}&&d_{\uparrow
}\left\vert \downarrow ,\,\tfrac{N-1}{2}\right\rangle =0, \\
d_{\downarrow }\left\vert \uparrow ,\,
\tfrac{N-1}{2}\right\rangle=0,\hspace{2.748cm}&&d_{\downarrow }\left\vert \downarrow ,\,\tfrac{N-1}{2}\right\rangle =\left\vert 0,\,\tfrac{N-1}{2}\right\rangle.
\end{eqnarray}%

If $N$ is even, the basis set of the subspace is $\left\vert 0,\,\frac{N}{2}\right\rangle $ and $\left\vert \uparrow
\downarrow ,\,\frac{N-2}{2}\right\rangle $. Note that
\begin{eqnarray}
H_{d}\left\vert 0,\,\tfrac{N}{2}%
\right\rangle  &=&\frac{1}{2}E_c(N-V_{g})^2 \left\vert 0,\,\tfrac{N}{2}\right\rangle +\Delta \left\vert \uparrow
\downarrow ,\,\tfrac{N-2}{2}\right\rangle,\\
H_d\left\vert \uparrow \downarrow ,\,\tfrac{N-2}{2}\right\rangle  &=&\Delta \left\vert 0,\,
\tfrac{N}{2}\right\rangle +\left[ 2\varepsilon _{0}+\frac{1}{2}E_c(
N-V_{g})^2 \right] \left\vert \uparrow \downarrow ,\,\tfrac{N-2}{2}\right\rangle,
\end{eqnarray}
with $\varepsilon_0\equiv\frac{1}{2}(\varepsilon_\uparrow+\varepsilon_\downarrow)$. The matrix form of $H_d$ is thus
\begin{eqnarray}
H_{d} =\begin{pmatrix} \frac{1}{2}E_c(N-V_{g})^2 & \Delta \\ \Delta
& 2\varepsilon _{0}+\frac{1}{2}E_c(N-V_{g})^2\end{pmatrix},
\end{eqnarray}
Its eigenenergies and eigenstates can be readily obtained as ($\nu=\pm$)
\begin{eqnarray}
E^\nu_N(V_g)&=&\varepsilon_0+\frac{1}{2}E_c(N-V_g)^2+\nu\sqrt{\Delta^2+\varepsilon_0^2},\\
|\Phi^\nu_N\rangle&=&l_{0\nu}\left|0,\,\tfrac{N}{2}\right\rangle+l_{2\nu}\left|\uparrow\downarrow,\,\tfrac{N-2}{2}\right\rangle,
\end{eqnarray}
where
\begin{equation}
l_{0\nu}\equiv\frac{c_\nu}{\sqrt{1+c_\nu^2}},\qquad\quad l_{2\nu}\equiv\frac{1}{\sqrt{1+c_\nu^2}},\qquad\quad c_\nu\equiv\frac{\nu\sqrt{\Delta^2+\varepsilon_0^2}-\varepsilon_0}{\Delta}.
\end{equation}
These coefficients satisfy $\sum_\nu l_{0\nu}^2=\sum_\nu l_{2\nu}^2=\sum_\nu \nu l_{0\nu}l_{2\bar\nu}=1$. Finally, the following formulas for $d_\uparrow$ and $d_\downarrow$ are useful for calculating their matrix elements which will be also used in the NRG procedure,
\begin{eqnarray}
d_{\uparrow }\left\vert 0,\,\tfrac{N}{2}%
\right\rangle=0,\hspace{1.5cm}&&d_{\uparrow }\left\vert \uparrow
\downarrow ,\,\tfrac{N-2}{2}\right\rangle
=\hspace{0.33cm} \left\vert \downarrow ,\,\tfrac{N-2}{2}\right\rangle,\\
d_{\downarrow }\left\vert 0,\,\tfrac{N}{2}%
\right\rangle=0,\hspace{1.5cm}&&d_{\downarrow }\left\vert \uparrow
\downarrow ,\,\tfrac{N-2}{2}\right\rangle
=-\left\vert \uparrow ,\,\tfrac{N-2}{2}\right\rangle.
\end{eqnarray}

\section{Appendix B: The Brillouin-Wigner perturbation theory}
\setcounter{equation}{0}
\renewcommand{\theequation}{B\arabic{equation}}
We consider now a system Hamiltonian $H$ that can be divided into two parts $H=H_0+H_T$. Suppose $H_0$ is exactly solvable and its eigenenergies and eigenstates are $E_i$ and $|i\rangle$, respectively, i.e.,
\begin{equation}
H_0|i\rangle=E_i|i\rangle,\qquad i=1,\,2,\,\cdots,\,m.
\end{equation}
Our purpose is to derive an effective Hamiltonian $H_{\text{eff}}$\ in the subspace
$span\left\{\,\left\vert 1\right\rangle,\,\left\vert 2\right\rangle,\,\cdots\cdots,\,\left\vert
m\right\rangle\,\right\}$, i.e., to project $H$ into the
subspace up to second order in $H_{T}$, following Ref.\,\cite{Fradkin2013}. The projection operator is $\hat{%
P}=\sum_{i=1}^{m}\left\vert i\right\rangle \left\langle i\right\vert $.
Supposing $\left\vert \Psi \right\rangle $ and $E$ are the eigenstate and
eigenenergy of the total Hamiltonian $H$, the effective Hamiltonian satisfies
\begin{equation}
H_{\text{eff}}\left\vert \psi \right\rangle =E\left\vert \psi \right\rangle,
\qquad\left\vert \psi \right\rangle \equiv \hat{P}\left\vert
\Psi \right\rangle.
\end{equation}
Starting from the Schr\"{o}dinger equation
\begin{equation}
H\left|\Psi\right\rangle=(H_0+H_T)\left|\Psi\right\rangle=E\left|\Psi\right\rangle,
\end{equation}
we formally get
\begin{eqnarray}
\left\vert \Psi \right\rangle &=&\left( E-H_{0}\right) ^{-1}H_{T}\left\vert
\Psi \right\rangle\nonumber \\
&=&\left( E-H_{0}\right) ^{-1}( 1-\hat{P}) H_{T}\left\vert
\Psi \right\rangle +\left( E-H_{0}\right) ^{-1}\sum_{i=1}^{m}\left\vert
i\right\rangle \left\langle i\right\vert H_{T}\left\vert \Psi \right\rangle\nonumber
\\
&=&G\left( E\right) H_{T}\left\vert \Psi \right\rangle +\sum_{i=1}^{m}\left(
E-E_{i}\right) ^{-1}\left\vert i\right\rangle \left\langle i\right\vert
H_{T}\left\vert \Psi \right\rangle,
\end{eqnarray}
where we have defined $G\left( E\right)
\equiv \left( E-H_{0}\right) ^{-1}( 1-\hat{P})$. On substituting
\begin{equation}
\langle i|H_T|\Psi\rangle=\langle i|(H-H_0)|\Psi\rangle=(E-E_i)\langle i|\Psi\rangle
\end{equation}
into Eq.\,(B4), one finds
\begin{equation}
|\Psi\rangle=G\left( E\right) H_{T}\left\vert \Psi \right\rangle +\left\vert \psi
\right\rangle.
\end{equation}
Equation (B6) can be iteratively solved, yielding
\begin{eqnarray}
|\Psi\rangle&=&\left\vert \psi \right\rangle +G\left( E\right) H_{T}\left\vert
\psi \right\rangle +\left[ G\left( E\right) H_{T}\right] ^{2}\left\vert
\psi \right\rangle +\left[ G\left( E\right) H_{T}\right] ^{3}\left\vert
\psi \right\rangle +\cdots\nonumber \\
&=&\sum_{n=0}^{\infty }\left[ G\left( E\right) H_{T}\right] ^{n}\left\vert
\psi \right\rangle.
\end{eqnarray}
Applying the operator $\hat{P}H_T$ to Eq.\,(B7) gives
\begin{equation}
\hat{P}H_{T}\left\vert \Psi
\right\rangle=\sum_{n=0}^{\infty }\hat{P}H_{T}\left[ G\left( E\right) H_{T}\right]
^{n}\left\vert \psi \right\rangle.
\end{equation}
The left-hand side of Eq.\,(B8) can be further calculated by using $\hat P^2=\hat P$ and $\hat PH_0=H_0\hat P$, as follows:
\begin{eqnarray}
\hat{P}H_{T}\left\vert \Psi
\right\rangle&=&\hat{P}H\left\vert \Psi
\right\rangle -\hat{P}^{2}H_{0}\left\vert \Psi \right\rangle\nonumber\\
&=&E\left\vert \psi \right\rangle -
\hat{P}H_{0}\left\vert \psi \right\rangle.
\end{eqnarray}
Substituting Eq.\,(B9) into Eq.\,(B8) and noting that $\hat P|\psi\rangle=|\psi\rangle$, we then arrive at the Schr\"{o}dinger equation in the subspace spanned by the eigenstates of $H_0$,
\begin{equation}
\left[ \hat{P}H_{0}\hat{P}+\sum_{n=0}^{\infty }\hat{P}H_{T}\left[
G\left( E\right) H_{T}\right] ^{n}\hat{P}\right] \left\vert \psi
\right\rangle=E\left\vert \psi \right\rangle.
\end{equation}
Therefore, the effective Hamiltonian is
\begin{eqnarray}
H_{\text{eff}} &=&\hat{P}H_{0}\hat{P}+\sum_{n=0}^{\infty }\hat{P}H_{T}%
\left[ G\left( E\right) H_{T}\right] ^{n}\hat{P}\nonumber\\
&\approx &\hat{P}H_{0}\hat{P}+\hat{P}H_{T}\hat{P}+\hat{P}%
H_{T}G\left( E\right) H_{T}\hat{P}\nonumber\\
&=&\hat{P}H_{0}\hat{P}+\hat{P}H_{T}\hat{P}+\hat{P}%
H_{T}\left( E-H_{0}\right) ^{-1}( 1-\hat{P}) H_{T}\hat{P}\nonumber\\
&=&\hat{P}H_{0}\hat{P}+\hat{P}H_{T}\hat{P}+\hat{P}%
H_{T}\left( E-H_{0}\right) ^{-1}H_{T}\hat{P}-\hat{P}H_{T}\left(
E-H_{0}\right) ^{-1}\hat{P}H_{T}\hat{P},
\end{eqnarray}
up to second order in $H_T$.

\section{Appendix C: Derivation of the spin and charge Kondo Hamiltonians}
\setcounter{equation}{0}
\renewcommand{\theequation}{C\arabic{equation}}
A superconducting quantum dot attached to the left ($\alpha=L$) and right ($\alpha=R$) normal metallic electrodes is described by the Hamiltonian
\begin{equation}
H=H_d+\sum_{k,\sigma,\alpha}\varepsilon_kC^\dagger_{k\sigma\alpha}C_{k\sigma\alpha}+\sum_{k,\sigma,\alpha}t_\alpha(C^\dagger_{k\sigma\alpha}d_\sigma+d^\dagger_\sigma C_{k\sigma\alpha}),
\end{equation}
with $H_d$ given by Eq.\,(A1). Here the left and right leads can be combined into a single effective lead coupled to the island. Specifically, we perform a canonical transformation $\left(t_0\equiv\sqrt{t_L^2+t_R^2}\right)$
\begin{eqnarray}
C_{k\sigma e}=\frac{t_L}{t_0}C_{k\sigma L}+\frac{t_R}{t_0}C_{k\sigma R},\qquad\quad &&C_{k\sigma o}=\frac{t_R}{t_0}C_{k\sigma L}-\frac{t_L}{t_0}C_{k\sigma R},\\
C_{k\sigma L}=\frac{t_L}{t_0}C_{k\sigma e}+\frac{t_R}{t_0}C_{k\sigma o},\qquad\quad &&C_{k\sigma R}=\frac{t_R}{t_0}C_{k\sigma e}-\frac{t_L}{t_0}C_{k\sigma o},
\end{eqnarray}
such that the resulting Hamiltonian contains only the $e$ mode of lead states, while the $o$ mode is completely decoupled. After dropping the index $e$ of the effective lead, the Hamiltonian reads
\begin{equation}
H=H_0+H_T,
\end{equation}
where $H_0=H_d+H_e$, $H_T=H_T^++H_T^-$, and
\begin{equation}
H_e=\sum_{k,\sigma}\varepsilon_kC^\dagger_{k\sigma}C_{k\sigma},\qquad H_T^+=\sum_{k,\sigma}t_0d^\dagger_\sigma C_{k\sigma},\qquad H_T^-=\sum_{k,\sigma}t_0C^\dagger_{k\sigma}d_\sigma.
\end{equation}
In the following, we shall derive, by using the Brillouin-Wigner perturbation theory up to second order in the tunneling term $H_T$, the spin and charge Kondo models from the Hamiltonian (C4). These Kondo effects take place when the gate voltage $V_g$ is tuned near an odd integer $2N_0+1$. Here $N_0$ is an arbitrary integer ensuring that the total number $N$ of electrons in the island always satisfies $N>0$.

{\it 1.\,The Spin Kondo Effect.---}We assume $\varepsilon _{\uparrow }=\varepsilon _{\downarrow }=\varepsilon
_{0}$ in the absence of external magnetic fields. When $\sqrt{\Delta
^{2}+\varepsilon _{0}^{2}}<\frac{E_c}{2}$ and
\begin{equation}
2N_{0}+\frac{1}{2}+\frac{1}{E_{c}}\sqrt{\Delta ^{2}+\varepsilon _{0}^{2}}%
<V_{g}<2N_{0}+\frac{3}{2}-\frac{1}{E_{c}}\sqrt{\Delta ^{2}+\varepsilon
_{0}^{2}},
\end{equation}
the two degenerate ground states of $H_{0}$ are
\begin{equation}
\big|\Phi^\uparrow_{2N_0+1}\big\rangle=\left\vert \uparrow ,\hspace{0.05cm}N_{0}\right\rangle,\qquad\quad\big|\Phi^\downarrow_{2N_0+1}\big\rangle=\left\vert
\downarrow,\hspace{0.05cm}N_{0}\right\rangle,
\end{equation}
with the ground-state energy
\begin{equation}
E^\sigma_{2N_0+1}(V_g)=\varepsilon _{0}+\frac{1}{2}E_c\left(2N_0+1-
V_{g}\right)^2\equiv E_0(V_g).
\end{equation}
The Fermi-sea state and energy of the effective lead are implicitly involved. The projection
operator for the subspace spanned by $\left\vert \uparrow ,\hspace{0.05cm}%
N_{0}\right\rangle $ and $\left\vert \downarrow ,\hspace{0.05cm}%
N_{0}\right\rangle \hspace{0.1cm}$ is
\begin{eqnarray}
\hat{P}&=&\sum_{\sigma }\left\vert \sigma ,\hspace{0.05cm}%
N_{0}\right\rangle \left\langle \sigma ,\hspace{0.05cm}N_{0}\right\vert\nonumber\\
&=&\sum_{\sigma }\left( \hat{N}_{\sigma }-\hat{N}_{\uparrow }\hat{N%
}_{\downarrow }\right) \left\vert N_{0};s\right\rangle \left\langle
N_{0};s\right\vert\nonumber\\
&=&\left( \hat{N}_{d}-2\hat{N}_{\uparrow }\hat{N%
}_{\downarrow }\right) \left\vert N_{0};s\right\rangle \left\langle
N_{0};s\right\vert,
\end{eqnarray}
where $\hat N_d=\sum_{\sigma=\uparrow,\downarrow}\hat N_\sigma$ and $N_\sigma=d_\sigma^\dagger d_\sigma$ are number operators of $d$ electrons.
Defining the spin operator $\mathbf{S}$ of the island as
\begin{equation}
S^{+}=d_{\uparrow }^{\dag }d_{\downarrow }\text{, \ \ \ }S^{-}=d_{\downarrow
}^{\dag }d_{\uparrow }\text{, \ \ \ }S^{z}=\frac{1}{2}\left( d_{\uparrow
}^{\dag }d_{\uparrow }-d_{\downarrow }^{\dag }d_{\downarrow }\right),
\end{equation}
we are going to derive an effective spin Kondo Hamiltonian $H_{\text{eff}}$\ in this $%
N=2N_{0}+1$ subspace. The first term of the effective Hamiltonian $H_{\text{eff}}$ [Eq.\,(B11)]
is
\begin{equation}
\hat{P}H_{0}\hat{P}=H_{e}\hat{P}+H_{d}\hat{P}=H_{e}\hat{P%
}+E_0(V_g)\hat{P}.
\end{equation}
The second and fourth terms of $H_{\text{eff}}$ [Eq.\,(B11)] are zero
\begin{equation}
\hat{P}H_{T}\hat{P}=0,\qquad\quad-\hat{P}H_{T}\left(
E-H_{0}\right) ^{-1}\hat{P}H_{T}\hat{P}=0.
\end{equation}

To derive the third term of $H_{\text{eff}}$ [Eq.\,(B11)], one needs to first calculate
\begin{eqnarray}
H_{T}^{-}\left\vert \sigma ,\hspace{0.05cm}N_{0}\right\rangle &=&\sum_{k,%
\hspace{0.05cm}\sigma ^{\prime }}t_0C_{k\sigma ^{\prime }}^{\dag }d_{\sigma ^{\prime }}\left\vert \sigma
,\hspace{0.05cm}N_{0}\right\rangle =\sum_{k}t_0C_{k\sigma}^{\dagger }\left\vert 0,\hspace{0.05cm}%
N_{0}\right\rangle, \\
H_{T}^{+}\left\vert \sigma ,\hspace{0.05cm}N_{0}\right\rangle &=&\sum_{k,%
\hspace{0.05cm}\sigma ^{\prime } }t_0d_{\sigma ^{\prime }}^{\dag }C_{k\sigma ^{\prime } }\left\vert \sigma
,\hspace{0.05cm}N_{0}\right\rangle =\sum_{k}\sigma
t_0C_{k\overline{\sigma } }\left\vert \uparrow \downarrow ,%
\hspace{0.05cm}N_{0}\right\rangle.
\end{eqnarray}
Since $\left\vert 0,\hspace{0.05cm}N_{0}\right\rangle $ and $\left\vert
\uparrow \downarrow ,\hspace{0.05cm}N_{0}\right\rangle $ are not eigenstates
of $H_{0}$, they should be expanded by eigenstates in the $N=2N_{0}$ and $%
N=2N_{0}+2$ subspaces, respectively,
\begin{eqnarray}
\left\vert 0,\hspace{0.05cm}N_{0}\right\rangle &=&l_{2-}\left\vert
\Phi_{2N_{0}}^+\right\rangle-l_{2+}\left\vert \Phi_{2N_{0}}^-\right\rangle, \\
\left\vert \uparrow \downarrow ,\hspace{0.05cm}N_{0}\right\rangle &=&-
l_{0-}\left\vert \Phi_{2N_{0}+2}^+\right\rangle+l_{0+}\left\vert
\Phi_{2N_{0}+2}^-\right\rangle.
\end{eqnarray}
One thus arrives at
\begin{eqnarray}
H_{T}^{-}\left\vert \sigma ,\hspace{0.05cm}N_{0}\right\rangle  &=&\sum_{k,\,\nu}\nu l_{2\bar\nu}t_0C_{k\sigma}^{\dag }\left\vert
\Phi_{2N_0}^\nu\right\rangle,\\
H_{T}^{+}\left\vert \sigma ,\hspace{0.05cm}N_{0}\right\rangle  &=&\sum_{k,\,\nu}\sigma\bar\nu l_{0\bar\nu} t_0C_{k\overline{\sigma}}\left\vert
\Phi_{2N_0+2}^\nu\right\rangle,
\end{eqnarray}
and
\begin{eqnarray}
\left\langle \sigma ,\hspace{0.05cm}N_{0}\right\vert H_{T}^{+} &=&\sum_{k,\,\nu}\nu l_{2\bar\nu}t_0\left\langle\Phi_{2N_0}^\nu\right\vert C_{k\sigma},\\
\left\langle \sigma ,\hspace{0.05cm}N_{0}\right\vert H_{T}^{-} &=&\sum_{k,\,\nu}\sigma\bar\nu l_{0\bar\nu}t_0\left\langle \Phi_{2N_0+2}^\nu\right\vert C_{k\overline{\sigma }}^{\dag },
\end{eqnarray}
\begin{eqnarray}
\big[ E_0(V_g)-H_{0}\big] ^{-1}H_{T}^{-}\left\vert \sigma ,\hspace{0.05cm}%
N_{0}\right\rangle  &=&\sum_{k,\,\nu}\frac{\nu l_{2\bar\nu}t_0}{
E_0(V_g)-E^\nu_{2N_0}(V_g)-\varepsilon_{k}}C_{k\sigma
}^{\dag }\left\vert \Phi^\nu_{2N_0}\right\rangle\nonumber\\
&=&\sum_{k,\,\nu}\frac{\nu l_{2\bar\nu}t_0}{
E_0(V_g)-E^\nu_{2N_0}(V_g)}C_{k\sigma
}^{\dag }\left\vert \Phi^\nu_{2N_0}\right\rangle,\\
\big[E_0(V_g)-H_{0}\big] ^{-1}H_{T}^{+}\left\vert \sigma ,\hspace{0.05cm}%
N_{0}\right\rangle  &=&\sum_{k,\,\nu}\frac{\sigma\bar\nu l_{0\bar\nu}t_0}{
E_0(V_g)-E^\nu_{2N_0+2}(V_g)+\varepsilon _{k}}C_{k\overline{\sigma
}}\left\vert \Phi^\nu_{2N_0+2}\right\rangle\nonumber\\
&=&\sum_{k,\,\nu}\frac{\sigma\bar\nu l_{0\bar\nu}t_0}{
E_0(V_g)-E^\nu_{2N_0+2}(V_g)}C_{k\overline{\sigma
}}\left\vert \Phi^\nu_{2N_0+2}\right\rangle.
\end{eqnarray}
The second lines of Eqs.\,(C21) and (C22) have applied the constraint $\varepsilon _{k}\simeq 0$ because at low temperatures only conduction electrons near the Fermi
energy ($\varepsilon _{F}=0$) in the effective lead can be exchange scattered by the superconducting quantum dot. A combination of Eqs.\,(C19)-(C22) yields
\begin{eqnarray}
\left\langle \sigma ,\hspace{0.05cm}%
N_{0}\right\vert H_{T}^{+}\big[E_0(V_g)-H_{0}\big]^{-1}H_{T}^{-}\left\vert
\sigma ^{\prime },\hspace{0.05cm}N_{0}\right\rangle&=&-\sum_{k,\,k^\prime}J_0C_{k\sigma}C^\dagger_{k^\prime\sigma^\prime},\\
\left\langle \sigma ,\hspace{0.05cm}
N_{0}\right\vert H_{T}^{-}\big[E_0(V_g)-H_{0}\big]^{-1}H_{T}^{+}\left\vert
\sigma^{\prime},\hspace{0.05cm}N_{0}\right\rangle&=&-\sum_{k,\,k^\prime}J_2\sigma\sigma^\prime C^\dagger_{k\overline\sigma}C_{k^\prime\overline{\sigma^\prime}}.
\end{eqnarray}
where
\begin{eqnarray}
J_{0}&\equiv&\sum_{\nu}\frac{l_{2\bar\nu}^{2}t_0^2}{E^{\nu}_{2N_0}(V_g)-E_0(V_g)}=\sum_{\nu}\frac{l_{2\bar\nu}^{2}t_0^2}{(V_g-2N_0-\frac{1}{2})E_c+\nu\sqrt{\Delta^2+\varepsilon_0^2}},\\
J_{2}&\equiv&\sum_{\nu}\frac{l_{0\bar\nu}^{2}t_0^2}{E^{\nu}_{2N_{0}+2}(V_{g})-E_0(V_g)}=\sum_{\nu}\frac{l_{0\bar\nu}^{2}t_0^2}{(2N_0+\frac{3}{2}-V_g)E_c+\nu\sqrt{\Delta^2+\varepsilon_0^2}}.
\end{eqnarray}
The third term of $H_{\text{eff}}$ [Eq.\,(B11)] can then be calculated as
\begin{eqnarray}
\hat{P}H_{T}\left( E-H_{0}\right) ^{-1}H_{T}\hat{P} &\approx &%
\hat{P}H_{T}\big[ E_0(V_g)-H_{0}\big] ^{-1}H_{T}\hat{P}\nonumber \\
&=&\hat{P}H_{T}^{+}\big[ E_0(V_g)-H_{0}\big] ^{-1}H_{T}^{-}\hat{P}+%
\hat{P}H_{T}^{-}\big[ E_0(V_g)-H_{0}\big] ^{-1}H_{T}^{+}\hat{P}\nonumber \\
&=&\sum_{\sigma ,\hspace{0.05cm}\sigma ^{\prime }}\left\vert \sigma ,\hspace{%
0.05cm}N_{0}\right\rangle \left\langle \sigma ,\hspace{0.05cm}%
N_{0}\right\vert H_{T}^{+}\big[ E_0(V_g)-H_{0}\big] ^{-1}H_{T}^{-}\left\vert
\sigma ^{\prime },\hspace{0.05cm}N_{0}\right\rangle \left\langle \sigma
^{\prime },\hspace{0.05cm}N_{0}\right\vert \nonumber \\
&&+\sum_{\sigma ,\hspace{0.05cm}\sigma ^{\prime }}\left\vert \sigma ,\hspace{%
0.05cm}N_{0}\right\rangle \left\langle \sigma ,\hspace{0.05cm}%
N_{0}\right\vert H_{T}^{-}\big[ E_0(V_g)-H_{0}\big] ^{-1}H_{T}^{+}\left\vert
\sigma ^{\prime },\hspace{0.05cm}N_{0}\right\rangle \left\langle \sigma
^{\prime },\hspace{0.05cm}N_{0}\right\vert \nonumber \\
&=&-\sum_{k,\,\sigma,\,k^\prime,\,\sigma^\prime }J_{0}C_{k\sigma
}C_{k^{\prime }\sigma ^{\prime }}^{\dag }\left\vert \sigma ,%
\hspace{0.05cm}N_{0}\right\rangle \left\langle \sigma ^{\prime },\hspace{%
0.05cm}N_{0}\right\vert\nonumber\\
&&-\sum_{k,\,\sigma,\,k^\prime,\,\sigma^\prime}J_{2}\sigma \sigma
^{\prime }C_{k\overline{\sigma } }^{\dag }C_{k^{\prime }\overline{%
\sigma ^{\prime }}}\left\vert \sigma ,\hspace{0.05cm}%
N_{0}\right\rangle \left\langle \sigma ^{\prime },\hspace{0.05cm}%
N_{0}\right\vert.
\end{eqnarray}
We continue to calculate
\begin{eqnarray}
-\sum_{k,\,\sigma,\,k^\prime,\,\sigma^\prime}J_{0}C_{k\sigma
}C_{k^{\prime }\sigma ^{\prime }}^{\dag }\left\vert \sigma ,%
\hspace{0.05cm}N_{0}\right\rangle \left\langle \sigma ^{\prime },\hspace{%
0.05cm}N_{0}\right\vert&=&-\sum_{k,\,\sigma,\,k^\prime,\,\sigma^\prime}J_{0}C_{k\sigma}C_{k^{\prime }\sigma ^{\prime }}^{\dag }\left(
d_{\sigma }^{\dag }d_{\sigma ^{\prime }}-\delta _{\sigma \sigma ^{\prime }}%
\hat{N}_{\uparrow }\hat{N}_{\downarrow }\right) \left\vert
N_{0};s\right\rangle \left\langle N_{0};s\right\vert\nonumber \\
&=&\sum_{k,\,k^\prime}J_{0}\left[
\begin{array}{c}
C_{k\uparrow }^{\dag }C_{k^{\prime }\downarrow}S^{-}+C_{k^{\prime }\downarrow}^{\dag }C_{k\uparrow}S^{+}\nonumber\\
+\left( C_{k\uparrow }^{\dag }C_{k^{\prime }\uparrow}-C_{k\downarrow }^{\dag }C_{k^{\prime }\downarrow }\right) S^{z}
\end{array}
\right] \left\vert N_{0};s\right\rangle \left\langle N_{0};s\right\vert\nonumber \\
&&+\sum_{k,\,k^\prime,\,\sigma}\frac{1}{2}J_{0}C_{k\sigma }^{\dag }C_{k^{\prime }\sigma}\hat{P}-\sum_{k}J_{0}\hat{P},
\end{eqnarray}
and
\begin{eqnarray}
-\sum_{k,\,\sigma,\,k^\prime,\,\sigma^\prime}J_{2}\sigma \sigma ^{\prime }C_{k%
\overline{\sigma } }^{\dag }C_{k^{\prime }\overline{\sigma ^{\prime }}}\left\vert \sigma ,\hspace{0.05cm}N_{0}\right\rangle
\left\langle \sigma ^{\prime },\hspace{0.05cm}N_{0}\right\vert&=&-\sum_{k,\,\sigma,\,k^\prime,\,\sigma^\prime}J_{2}\sigma \sigma ^{\prime }C_{k\overline{\sigma }%
 }^{\dag }C_{k^{\prime }\overline{\sigma ^{\prime }}}\left( d_{\sigma }^{\dag }d_{\sigma ^{\prime }}-\delta _{\sigma \sigma
^{\prime }}\hat{N}_{\uparrow }\hat{N}_{\downarrow }\right)
\left\vert N_{0};s\right\rangle \left\langle N_{0};s\right\vert\nonumber\\
&=&\sum_{k,\,k^{\prime}}J_{2}\left[
\begin{array}{c}
C_{k\uparrow }^{\dag }C_{k^{\prime }\downarrow}S^{-}+C_{k^{\prime }\downarrow}^{\dag }C_{k\uparrow
}S^{+}\nonumber\\
+\left( C_{k\uparrow}^{\dag }C_{k^{\prime }\uparrow}-C_{k\downarrow}^{\dag }C_{k^{\prime }\downarrow }\right) S^{z}
\end{array}
\right] \left\vert N_{0};s\right\rangle \left\langle N_{0};s\right\vert\nonumber \\
&&-\sum_{k,\,k^\prime,\,\sigma}\frac{1}{2}J_{2}C_{k\sigma}^{\dag }C_{k^{\prime }\sigma}\hat{P}.
\end{eqnarray}

Combining the above two equations with the lead Hamiltonian $H_{e}\hat{P}
$ and discarding the constant terms, the effective Hamitonian reads
\begin{eqnarray}
H_{\text{eff}}&=&\sum_{k,\,k^{\prime }}\left( J_{0}+J_{2}\right) \left[
C_{k\uparrow}^{\dag }C_{k^{\prime }\downarrow
}S^{-}+C_{k^{\prime }\downarrow }^{\dag }C_{k\uparrow
}S^{+}+\left( C_{k\uparrow}^{\dag }C_{k^{\prime }\uparrow
}-C_{k\downarrow}^{\dag }C_{k^{\prime }\downarrow
}\right) S^{z}
\right] \left\vert N_{0};s\right\rangle \left\langle N_{0};s\right\vert\nonumber\\
&&+\sum_{k,\,k^\prime,\,\sigma}\frac{1}{2}\left( J_{0}-J_{2}\right)
C_{k\sigma}^{\dag }C_{k^{\prime }\sigma}\hat{P}\ +\ \sum_{k,\,\sigma}\varepsilon
_{k}C_{k\sigma}^{\dag }C_{k\sigma}\hat{P}.
\end{eqnarray}
This effective Hamiltonian describes the isotropic spin Kondo effect in the $%
N=2N_{0}+1$ subspace. Note that for $\varepsilon_0=0$ and $V_g=2N_0+1$, one has $J_0=J_2$, which eliminates the potential scattering term in $H_{\text{eff}}$. The first line of Eq.\,(C30) gives Eq.\,(4) in the main text.

{\it 2.\,The Charge Kondo Effect.---} For $\sqrt{\Delta ^{2}+\varepsilon _{0}^{2}}>\frac{E_c}{2}$ and $%
V_{g}=2N_{0}+1\equiv V_{g0}$, the two degenerate ground states and the
ground-state energy of $H_{0}$ are
\begin{eqnarray}
\left\vert\Phi^-_{2N_0}\right\rangle&=&l_{0-}\left\vert 0,\hspace{0.05cm}%
N_{0}\right\rangle +l_{2-}\left\vert \uparrow \downarrow ,\hspace{0.05cm}%
N_{0}-1\right\rangle, \\
\left\vert\Phi^-_{2N_0+2}\right\rangle&=&l_{0-}\left\vert 0,\hspace{0.05cm}%
N_{0}+1\right\rangle +l_{2-}\left\vert \uparrow \downarrow ,\hspace{0.05cm}%
N_{0}\right\rangle, \\
E^-_{2N_0}(V_{g0})&=&E^-_{2N_0+2}(V_{g0})\ =\ \varepsilon_{0}+\frac{1}{2}E_{c}-\sqrt{\Delta ^{2}+\varepsilon
_{0}^{2}}\ \equiv\ E_0.
\end{eqnarray}
The Fermi-sea state and energy of the effective electrode are implicitly involved. By rewriting the two ground
states into a more succinct form ($\eta=0,1$)
\begin{equation}
\big\vert\Phi^-_{2N_{0}+2\eta}\big\rangle=l_{0-}\left\vert 0,\hspace{0.05cm}%
N_{0}+\eta \right\rangle +l_{2-}\left\vert \uparrow \downarrow ,\hspace{0.05cm%
}N_{0}+\eta -1\right\rangle,
\end{equation}
the projection operator onto the ground-state subspace is thus
\begin{equation}
\hat{P}=\sum_{\eta}\big\vert\Phi^-_{2N_{0}+2\eta}\big\rangle\big\langle\Phi^-_{2N_{0}+2\eta}
\big\vert.
\end{equation}
Moreover, the states $\left\vert\Phi^-_{2N_0}\right\rangle$ and $\left\vert\Phi^-_{2N_0+2}\right\rangle$ represent two different charge states of the superconducting quantum dot, based on which, we can define its pseudospin operator $\mathbf{Q}$ as
\begin{eqnarray}
Q^{+}&=&\left\vert\Phi^-_{2N_{0}+2}\right\rangle\left\langle\Phi^-_{2N_0}\right\vert,\\
Q^{-}&=&\left\vert\Phi^-_{2N_0}\right\rangle\left\langle
\Phi^-_{2N_{0}+2}\right\vert,\\
Q^{z}&=&\frac{1}{2}\Big( \left\vert
\Phi^-_{2N_{0}+2}\right\rangle\left\langle\Phi^-_{2N_{0}+2}\right\vert-\left\vert
\Phi^-_{2N_0}\right\rangle\left\langle\Phi^-_{2N_0}\right\vert\Big).
\end{eqnarray}

In the subspace spanned by $\left\vert\Phi^-_{2N_0}\right\rangle$ and $
\left\vert\Phi^-_{2N_{0}+2}\right\rangle$, the first term of the effective
Hamiltonian $H_{\text{eff}}$ [Eq.\,(B11)] is
\begin{equation}
\hat{P}H_{0}\hat{P}=H_{e}\hat{P}+H_d\hat{P}=H_{e}\hat{P}+E_{0}\hat{P}.
\end{equation}
The second and fourth terms of $H_{\text{eff}}$ [Eq.\,(B11)] are zero
\begin{equation}
\hat{P}H_{T}\hat{P}=0,\qquad\quad-\hat{P}H_{T}\left(
E-H_{0}\right) ^{-1}\hat{P}H_{T}\hat{P}=0.
\end{equation}
In order to derive the third term of $H_{\text{eff}}$ [Eq.\,(B11)], one needs to first
calculate
\begin{eqnarray}
H_{T}^{-}\big\vert\Phi^-_{2N_{0}+2\eta} \big\rangle &=&\sum_{k,\,\sigma }l_{2-}t_0C_{k\sigma}^{\dag
}d_{\sigma }\left\vert \uparrow \downarrow ,\hspace{0.05cm}N_{0}+\eta
-1\right\rangle\nonumber\\
&=&\sum_{k,\,\sigma}\sigma l_{2-}t_0C_{k\sigma}^{\dag }\left\vert \overline{%
\sigma },\hspace{0.05cm}N_{0}+\eta -1\right\rangle,\\
H_{T}^{+}\big\vert\Phi^-_{2N_{0}+2\eta}\big\rangle &=&\sum_{k,\,\sigma}l_{0-}t_0d_{\sigma }^{\dag
}C_{k\sigma}\left\vert 0,\hspace{0.05cm}N_{0}+\eta \right\rangle\nonumber\\
&=&-\sum_{k,\,\sigma}l_{0-}t_0C_{k\sigma}\left\vert \sigma ,\hspace{0.05cm}N_{0}+\eta
\right\rangle,
\end{eqnarray}
and then
\begin{eqnarray}
\big\langle\Phi^-_{2N_{0}+2\eta}\big\vert H_{T}^{+}&=&\sum_{k,\,\sigma}\sigma l_{2-}t_0\left\langle
\overline{\sigma },\hspace{0.05cm}N_{0}+\eta -1\right\vert C_{k\sigma},\\
\big\langle\Phi^-_{2N_{0}+2\eta} \big\vert H_{T}^{-} &=&-\sum_{k,\,\sigma}l_{0-}t_0\left\langle \sigma ,%
\hspace{0.05cm}N_{0}+\eta \right\vert C_{k\sigma}^{\dag },
\end{eqnarray}
\begin{eqnarray}
\left( E_0-H_{0}\right)^{-1}H_{T}^{-}\big\vert\Phi^-_{2N_{0}+2\eta}
\big\rangle&=&\sum_{k,\,\sigma}
\frac{\sigma l_{2-}t_0}{E_0-E^{\overline\sigma}_{2N_{0}+2\eta -1}\left(
V_{g0}\right) -\varepsilon _{k}}C_{k\sigma}^{\dag }\left\vert
\overline{\sigma },\hspace{0.05cm}N_{0}+\eta -1\right\rangle\nonumber \\
&=&\sum_{k,\,\sigma}\frac{\sigma
l_{2-}t_0}{E_0-E^{0}_{2N_{0}+2\eta -1}\left( V_{g0}\right) }%
C_{k\sigma}^{\dag }\left\vert \overline{\sigma },\hspace{0.05cm}%
N_{0}+\eta -1\right\rangle, \\
\left( E_0-H_{0}\right) ^{-1}H_{T}^{+}\big\vert\Phi^-_{2N_{0}+2\eta}
\big\rangle&=&-\sum_{k,\,\sigma}
\frac{l_{0-}t_0}{E_0-E^{\sigma}_{2N_{0}+2\eta +1}\left( V_{g0}\right)
+\varepsilon _{k}}C_{k\sigma}\left\vert \sigma ,\hspace{0.05cm}%
N_{0}+\eta \right\rangle\nonumber \\
&=&-\sum_{k,\,\sigma}\frac{
l_{0-}t_0}{E_0-E^{0}_{2N_{0}+2\eta +1}\left( V_{g0}\right) }%
C_{k\sigma}\left\vert \sigma ,\hspace{0.05cm}N_{0}+\eta \right\rangle.
\end{eqnarray}
The second lines of Eqs.\,(C45) and (C46) have used $\varepsilon_{k}\simeq 0$ and for an odd number $N$ of electrons in the island,
\begin{equation}
E^{\sigma}_{N}\left( V_{g}\right)=\varepsilon _{0}+Q\left(N-
V_{g}\right) \equiv E^{0}_N\left( V_{g}\right),
\end{equation}
in the absence of external magnetic fields. Combining Eqs.\,(C43)-(C46) yields
\begin{eqnarray}
\big\langle\Phi^-_{2N_{0}+2\eta} \big\vert
H_{T}^{-}\left( E_0-H_{0}\right) ^{-1}H_{T}^{-}\big\vert\Phi^-_{2N_{0}+2\eta
^{\prime }}\big\rangle&=&\delta_{\eta0}\delta_{\eta^\prime1}J_1\sum_{k,\,k^\prime}C^\dagger_{k\uparrow}C^\dagger_{k^\prime\downarrow},\\
\big\langle\Phi^-_{2N_{0}+2\eta} \big\vert H_{T}^{+}\left( E_0-H_{0}\right) ^{-1}H_{T}^{+}\big\vert\Phi^-_{2N_{0}+2\eta
^{\prime }}\big\rangle&=&\delta_{\eta1}\delta_{\eta^\prime0}J_1\sum_{k,\,k^\prime}C_{k^\prime\downarrow}C_{k\uparrow},\\
\big\langle\Phi^-_{2N_{0}+2\eta} \big\vert H_{T}^{-}\left( E_0-H_{0}\right) ^{-1}H_{T}^{+}\big\vert\Phi^-_{2N_{0}+2\eta
^{\prime }}\big\rangle&=&\delta_{\eta\eta^\prime}\frac{l_{0-}}{2l_{2-}}J_{2\eta+1}\sum_{k,\,k^\prime,\,\sigma}C^\dagger_{k\sigma}C_{k^\prime\sigma},\\
\big\langle\Phi^-_{2N_{0}+2\eta} \big\vert H_{T}^{+}\left( E_0-H_{0}\right) ^{-1}H_{T}^{-}\big\vert\Phi^-_{2N_{0}+2\eta
^{\prime }}\big\rangle&=&\delta_{\eta\eta^\prime}\frac{l_{2-}}{2l_{0-}}J_{2\eta-1}\sum_{k,\,k^\prime,\,\sigma}\left(\delta_{kk^\prime}-C^\dagger_{k\sigma}C_{k^\prime\sigma}\right),
\end{eqnarray}
where ($y=-1,\,1,\,3$)
\begin{equation}
J_y\equiv\frac{2l_{0-}l_{2-}t^2_0}{E_0-E^0_{2N_0+y}(V_{g0})}=\frac{4l_{0-}l_{2-}t^2_0}{(2y-y^2)E_c-2\sqrt{\Delta^2+\varepsilon_0^2}}.
\end{equation}
Thus, the third term of $H_{\text{eff}}$ [Eq.\,(B11)] can be calculated
as follows
\begin{eqnarray}
\hat{P}H_{T}\left( E-H_{0}\right) ^{-1}H_{T}\hat{P}&\approx &
\hat{P}H_{T}\left( E_0-H_{0}\right) ^{-1}H_{T}\hat{P}\nonumber \\
&=&\sum_{\eta,\,\eta^\prime}\big\vert
\Phi^-_{2N_{0}+2\eta}\big\rangle\big\langle\Phi^-_{2N_{0}+2\eta}\big\vert H_{T}^{-}\left( E_{0}-H_{0}\right) ^{-1}H_{T}^{-}\big\vert\Phi^-_{2N_{0}+2\eta
^\prime}\big\rangle\big\langle\Phi^-_{2N_{0}+2\eta ^\prime}\big\vert\nonumber\\
&&+\sum_{\eta,\,\eta^\prime}\big\vert
\Phi^-_{2N_{0}+2\eta}\big\rangle\big\langle\Phi^-_{2N_{0}+2\eta}\big\vert H_{T}^{+}\left( E_{0}-H_{0}\right) ^{-1}H_{T}^{+}\big\vert\Phi^-_{2N_{0}+2\eta
^\prime}\big\rangle\big\langle\Phi^-_{2N_{0}+2\eta ^\prime}\big\vert\nonumber\\
&&+\sum_{\eta,\,\eta^\prime}\big\vert
\Phi^-_{2N_{0}+2\eta}\big\rangle\big\langle\Phi^-_{2N_{0}+2\eta}\big\vert H_{T}^{-}\left( E_{0}-H_{0}\right) ^{-1}H_{T}^{+}\big\vert\Phi^-_{2N_{0}+2\eta
^\prime}\big\rangle\big\langle\Phi^-_{2N_{0}+2\eta ^\prime}\big\vert\nonumber\\
&&+\sum_{\eta,\,\eta^\prime}\big\vert
\Phi^-_{2N_{0}+2\eta}\big\rangle\big\langle\Phi^-_{2N_{0}+2\eta}\big\vert H_{T}^{+}\left( E_{0}-H_{0}\right) ^{-1}H_{T}^{-}\big\vert\Phi^-_{2N_{0}+2\eta
^\prime}\big\rangle\big\langle\Phi^-_{2N_{0}+2\eta ^\prime}\big\vert\nonumber\\
&=&\sum_{k,\,k^\prime}J_{1}C_{k\uparrow}^{\dag }C_{k^{\prime }\downarrow}^{\dag }Q^-+\sum_{k,\,k^\prime}J_{1}C_{k^{\prime }\downarrow}C_{k\uparrow}Q^+\nonumber\\
&&+\sum_{k,\,k^\prime,\,\sigma}\sum_\eta\left( \frac{l_{0-}}{2l_{2-}}J_{2\eta +1}-\frac{l_{2-}}{2l_{0-}}J_{2\eta -1}\right)
\left(C_{k\sigma }^{\dag }C_{k^{\prime }\sigma}-\frac{1}{2}\delta _{kk^{\prime }}\right)
\big\vert\Phi^-_{2N_{0}+2\eta}\big\rangle\big\langle\Phi^-_{2N_{0}+2\eta}
\big\vert\nonumber\\
&&+\sum_{k,\,k^\prime}\sum_{\eta}\left( \frac{l_{0-}}{2l_{2-}}J_{2\eta +1}+\frac{l_{2-}}{2l_{0-}}J_{2\eta -1}\right)\delta_{kk^\prime}\big\vert
\Phi^-_{2N_{0}+2\eta}\big\rangle\big\langle\Phi^-_{2N_{0}+2\eta}\big\vert.
\end{eqnarray}
We proceed to recast the last two lines of Eq.\,(C53) in terms of $Q^z$ and $\hat P$,
\begin{eqnarray}
&&\sum_{k,\,k^\prime,\,\sigma}\sum_\eta\left( \frac{l_{0-}}{2l_{2-}}J_{2\eta +1}-\frac{l_{2-}}{2l_{0-}}J_{2\eta -1}\right)
\left(C_{k\sigma }^{\dag }C_{k^{\prime }\sigma}-\frac{1}{2}\delta _{kk^{\prime }}\right)
\big\vert\Phi^-_{2N_{0}+2\eta}\big\rangle\big\langle\Phi^-_{2N_{0}+2\eta}
\big\vert\nonumber\\
&=&\sum_{k,\,k^\prime,\,\sigma}\left[\frac{l_{2-}}{2l_{0-}}J_{-1}
-\left(\frac{l_{2-}}{2l_{0-}}+\frac{l_{0-}}{2l_{2-}}\right)J_1
+\frac{l_{0-}}{2l_{2-}}J_3\right]
\left(C_{k\sigma }^{\dag }C_{k^{\prime }\sigma}-\frac{1}{2}\delta_{kk^{\prime }}\right)Q^z\nonumber\\
&&+\ \sum_{k,\,k^\prime,\,\sigma}\left[-\frac{l_{2-}}{4l_{0-}}J_{-1}
-\left(\frac{l_{2-}}{4l_{0-}}-\frac{l_{0-}}{4l_{2-}}\right)J_1
+\frac{l_{0-}}{4l_{2-}}J_3\right]
\left(C_{k\sigma }^{\dag }C_{k^{\prime }\sigma}-\frac{1}{2}\delta_{kk^{\prime }}\right)\hat P,
\end{eqnarray}
and
\begin{eqnarray}
&&\sum_{k,\,k^\prime}\sum_{\eta}\left( \frac{l_{0-}}{2l_{2-}}J_{2\eta +1}+\frac{l_{2-}}{2l_{0-}}J_{2\eta -1}\right)\delta_{kk^\prime}\big\vert
\Phi^-_{2N_{0}+2\eta}\big\rangle\big\langle\Phi^-_{2N_{0}+2\eta}\big\vert\nonumber\\
&=&\sum_{k,\,k^\prime}\left[-\frac{l_{2-}}{2l_{0-}}J_{-1}+\left(\frac{l_{2-}}{2l_{0-}}
-\frac{l_{0-}}{2l_{2-}}\right)J_{1}+\frac{l_{0-}}{2l_{2-}}J_3\right]\delta_{kk^\prime}Q^z\nonumber\\
&&+\ \,\sum_{k,\,k^\prime}\left[\frac{l_{2-}}{4l_{0-}}J_{-1}+\left(\frac{l_{2-}}{4l_{0-}}
+\frac{l_{0-}}{4l_{2-}}\right)J_{1}+\frac{l_{0-}}{4l_{2-}}J_3\right]\delta_{kk^\prime}\hat P.
\end{eqnarray}

Collecting all contributing terms and discarding the constant ones, we finally
obtain the effective Hamiltonian
\begin{eqnarray}
H_{\text{eff}} &=&\sum_{k,\,k^\prime}J_{\pm}\left(C_{k\uparrow}^{\dag }C_{k^{\prime }\downarrow}^{\dag }Q^{-}
+C_{k^{\prime }\downarrow}C_{k\uparrow}Q^{+}\right)\,
+\sum_{k,\,k^\prime,\,\sigma}J_{z}\left(C_{k\sigma}^{\dag }C_{k^{\prime }\sigma}
-\frac{1}{2}\delta _{kk^{\prime }}\right) Q^{z}\nonumber\\
&&+\ \tilde{B}Q^{z}+\sum_{k,\,k^\prime,\,\sigma}VC_{k\sigma}^{\dag }C_{k^{\prime }\sigma}\hat{%
P}\,+\,\sum_{k,\,\sigma}\varepsilon_{k}C_{k\sigma}^{\dag }C_{k\sigma}\hat{P},
\end{eqnarray}
where
\begin{eqnarray}
J_{\pm }&\equiv&J_1\,=\,\frac{4l_{0-}l_{2-}t^2_0}{E_c-2\sqrt{\Delta^2+\varepsilon^2_0}},\\
J_{z}&\equiv&\frac{l_{2-}}{2l_{0-}}J_{-1}
-\left(\frac{l_{2-}}{2l_{0-}}+\frac{l_{0-}}{2l_{2-}}\right)J_1
+\frac{l_{0-}}{2l_{2-}}J_3\nonumber\\
&=&\frac{2t_0^2}{2\sqrt{\Delta^2+\varepsilon_0^2}-E_c}-\frac{2t_0^2}{3E_c+2\sqrt{\Delta^2+\varepsilon_0^2}},\\
\tilde{B} &\equiv&\sum_{k,\,k^\prime}\left[-\frac{l_{2-}}{2l_{0-}}J_{-1}+\left(\frac{l_{2-}}{2l_{0-}}
-\frac{l_{0-}}{2l_{2-}}\right)J_{1}+\frac{l_{0-}}{2l_{2-}}J_3\right]\delta_{kk^\prime}\nonumber\\
&=&\left(l_{0-}^2-l_{2-}^2\right)\sum_{k,\,k^\prime}J_z\delta_{kk^\prime},\\
V&\equiv&-\frac{l_{2-}}{4l_{0-}}J_{-1}
-\left(\frac{l_{2-}}{4l_{0-}}-\frac{l_{0-}}{4l_{2-}}\right)J_1
+\frac{l_{0-}}{4l_{2-}}J_3\nonumber\\
&=&\left(l^2_{2-}-l^2_{0-}\right)\left(\frac{t_0^2}{2\sqrt{\Delta^2+\varepsilon_0^2}-E_c}+\frac{t_0^2}{3E_c+2\sqrt{\Delta^2+\varepsilon_0^2}}\right).
\end{eqnarray}
This effective Hamiltonian describes the charge Kondo effect in the subspace
spanned by the degenerate states $\left\vert\Phi^-_{2N_{0}}\right\rangle$ and $
\left\vert\Phi^-_{2N_{0}+2}\right\rangle$ at $V_{g}=V_{g0}$. Some comments are necessary. (i) The
charge Kondo effect is anisotropic because the transverse exchange
interaction $J_{\pm}$ and the
longitudinal one $J_{z}$ are
different. (ii) For an isolated island at $V_{g}=V_{g0}$, the pseudospin-up
state $\left\vert\Phi^-_{2N_{0}+2}\right\rangle$ and the pseudospin-down state $%
\left\vert\Phi^-_{2N_{0}}\right\rangle$ are always degenerate for arbitrary
values of the $d$-level energy $\varepsilon _{0}$. (iii) The presence of dot-lead
coupling $t_0$ can induce an
effective local magnetic field $\tilde{B}$, which lifts out the degeneracy and suppresses the charge Kondo effect. However, the exchange field $\tilde B$ vanishes at $\varepsilon_0=0$ because of $l_{0-}^{2}=l_{2-}^{2}$. (iv) The trivial scattering potential $V$ also vanishes at $\varepsilon_0=0$. The first three terms of Eq.\,(C56) give Eq.\,(6) in the main text.

\end{appendix}
\end{widetext}

\end{document}